\documentclass[traditabstract]{aa}
\bibpunct{(}{)}{;}{a}{}{,} 
\usepackage{graphicx}
\usepackage{mathrsfs}
\usepackage[varg]{txfonts}
\usepackage{color}
\usepackage{stfloats}
\usepackage[switch,pagewise]{lineno}

\usepackage{amstext}

\begin{document}

\title{Geometry and kinematics of the broad emission line region in the lensed quasar Q2237+0305}
\author{D. Hutsem\'ekers\inst{1,}\thanks{Research Director F.R.S.-FNRS},
        D. Sluse\inst{1}
        }
\institute{
    Institut d'Astrophysique et de G\'eophysique,
    Universit\'e de Li\`ege, All\'ee du 6 Ao\^ut 19c, B5c,
    4000 Li\`ege, Belgium
    }
\date{Received ; accepted: }
\titlerunning{Microlensing of the broad line region in Q2237+0305} 
\authorrunning{D. Hutsem\'ekers and D. Sluse}
\abstract{
Line profile distortions are commonly observed in gravitationally lensed quasar spectra. These distortions are caused by microlensing from the stars in the lensing galaxy, which produce differential magnification of spatially and kinematically separated parts of the broad line region (BLR). The quasi-simultaneous visible and near-infrared spectroscopy of the lensed quasar Q2237+0305 reveals strong microlensing-induced line deformations in the high-ionization \ion{C}{iv} $\lambda 1549 \AA$ and the low-ionization H$\alpha$ emission lines. We use this effect to constrain the BLR size, geometry, and kinematics in Q2237+0305. For this purpose, we modeled the deformation of the emission lines for three representative BLR models: a Keplerian disk, an equatorial wind, and a biconical polar wind. We considered various inclinations with respect to the line of sight. We find that the observed microlensing effect, characterized by a set of four indices, can only be reproduced by a subsample of the considered BLR models. The microlensing analysis favors a Keplerian disk model for the regions emitting the \ion{C}{iv} and the H$\alpha$ emission lines. A polar wind model remains possible for the \ion{C}{iv} BLR, although it is less likely. The equatorial wind model is totally excluded. A preferred inclination of the BLR of 40$\degr$ is found, in agreement with expectations for a type 1 AGN and past constraints on the accretion disk inclination. The half-light radius of the BLR is $ r_{\rm 1/2} \simeq$ 47$\pm$19 light-days, with no significant difference between the \ion{C}{iv} and H$\alpha$ BLRs. The size of the \ion{C}{iv} BLR agrees with the radius-luminosity relation derived from reverberation mapping, while the size of the Balmer line BLR is one order of magnitude smaller, possibly revealing different quasar properties at high luminosities and high accretion rates.
}
\maketitle
%
%
%
\section{Introduction}
\label{sec:intro}

The  geometry and kinematics of the broad line region (BLR) in active galactic nuclei (AGNs) has been essentially investigated through reverberation mapping \citep{1982BlandfordMcKee,1993Peterson}, a technique that takes advantage of the variability of the ionizing continuum to measure the response time of the emission lines, which is directly related to the size of the BLR. Time lags as a function of the velocity across the H$\beta$ emission line profile have been measured in a dozen AGNs, providing information on the geometry and kinematics of the BLR. The geometry is generally a thick disk viewed close to face-on, while various kinematical signatures were found in the different objects, mostly virialized motions, inflows, or outflows \citep[e.g.,][]{2009Bentz, 2010Bentz, 2010Denney, 2014Pancoast, 2016Du, 2017Grier, 2018Xiao,2019Zhang}. Recently, the advent of long-baseline near-infrared interferometry made it possible to spatially resolve the P$\alpha$ and Br$\gamma$ BLRs in two AGNs, revealing a thick disk dominated by Keplerian rotation \citep{2018Gravity,2021Gravity}. These observations agree with the results from reverberation mapping.

Several studies have shown that line profile distortions are commonly observed in gravitationally lensed quasar spectra \citep{2004Richards, 2005Wayth, 2007Sluse, 2011Sluse, 2012Sluse, 2011ODowd, 2013Guerras, 2014Braibant, 2016Braibant, 2017Motta, 2018Fian, 2020Popovic}. These line profile deformations are most often detected as red/blue or wings/core distortions \citep{2012Sluse}. Simulations have shown that reverberation is unable to reproduce the line deformations observed in most systems when the delay is shorter than 40 days. In these cases, the line deformations can be  attributed to microlensing, that is, the differential magnification of spatially and kinematically separated subregions of the BLR. Microlensing-induced line profile distortions can thus provide independent constraints on the BLR structure. 

Possible effects of microlensing on broad emission lines have been theoretically investigated by several authors \citep{1988Nemiroff, 1990SchneiderWambsganss, 1994Hutsemekers, 2001Popovic, 2002Abajas, 2007Abajas, 2004LewisIbata, 2011ODowd, 2011Garsden, 2014Simic}. In particular, \citet{2017Braibant} have recently computed the effect of gravitational microlensing on quasar broad emission line profiles and their underlying continuum, considering representative BLR models and microlensing magnification maps. To analyze the large amount of simulated line profiles, the effects of microlensing have been summarized using four observables:  $\mu^{cont}$, the magnification of the continuum at the wavelength of the emission line, $\mu^{BLR}$, the total magnification of the broad emission line, and two indices sensitive to red/blue and wings/core line profile distortions, denoted $RBI$ and $WCI$, respectively.  $WCI,RBI$ diagrams that can serve as diagnostic diagrams to distinguish the different BLR models were then built. Furthermore, \citet{2019Hutsemekers} have developed a Bayesian probabilistic scheme to identify the models that best reproduce the observed line profile deformations. By comparing the microlensing-induced distortions of the H$\alpha$ emission line observed in the lensed quasar HE0435$-$1223 with a large number of simulations, the authors concluded that BLR flattened geometries, such as a Keplerian disk or an equatorial wind, reproduce the observations best.

The quadruply lensed quasar Q2237+0305 at $z$ = 1.695 \citep{1985Huchra}, also known as the Einstein cross, has been known for years to be a privileged laboratory for microlensing studies. The very short time delays of about one day \citep[e.g.,][]{2006Vakulik} allow one to robustly interpret the spectral differences observed between the lensed images in terms of microlensing. \citet{2011Sluse} measured a size of the \ion{C}{iv} emitting region that agrees with the radius-luminosity relation derived from reverberation mapping. These authors also found that the broadest components of the \ion{C}{iv} and \ion{C}{iii}] lines are more strongly microlensed than the line cores, suggesting that the highest velocity parts of the profiles come from the most compact regions of the BLR. \citet{2011ODowd} found that the differential microlensing signature observed in the \ion{C}{iii}] emission line favors a gravitationally dominated BLR kinematics. \citet{2016Braibant} used a combination of optical and near-infrared spectra of the four images and reported the existence in image A of a large-amplitude microlensing effect that distorts the high-ionization \ion{C}{iv} $\lambda 1549 \AA$ line and the low-ionization H$\alpha$ line in different ways.  Making use of a dedicated disentangling method, they argued that the differential magnification of the blue and red wings of H$\alpha$ favors a flattened, rotating, low-ionization region, whereas the roughly symmetric microlensing effect measured in \ion{C}{iv} can be reproduced assuming that the high-ionization emission line is formed in a polar wind.

In order to place these results in a more quantitative framework, we now compare these observations with simulations for a set of simple but representative BLR models. We first characterize  the microlensing effect in the  \ion{C}{iv} and H$\alpha$ lines of Q2237+0305 by measuring the four indices $\mu^{cont}$, $\mu^{BLR}$, $RBI,$ and $WCI$ (Sect.~\ref{sec:measurements}). These indices are then compared with those obtained from the simulations, with the aim to constrain the BLR properties (Sect.~\ref{sec:modeling}). Conclusions form the last section.

\section{Measurement of magnification and distortion indices}
\label{sec:measurements}

\begin{figure*}
\resizebox{\hsize}{!}{\includegraphics*[trim={0 0 0 0 mm},clip]{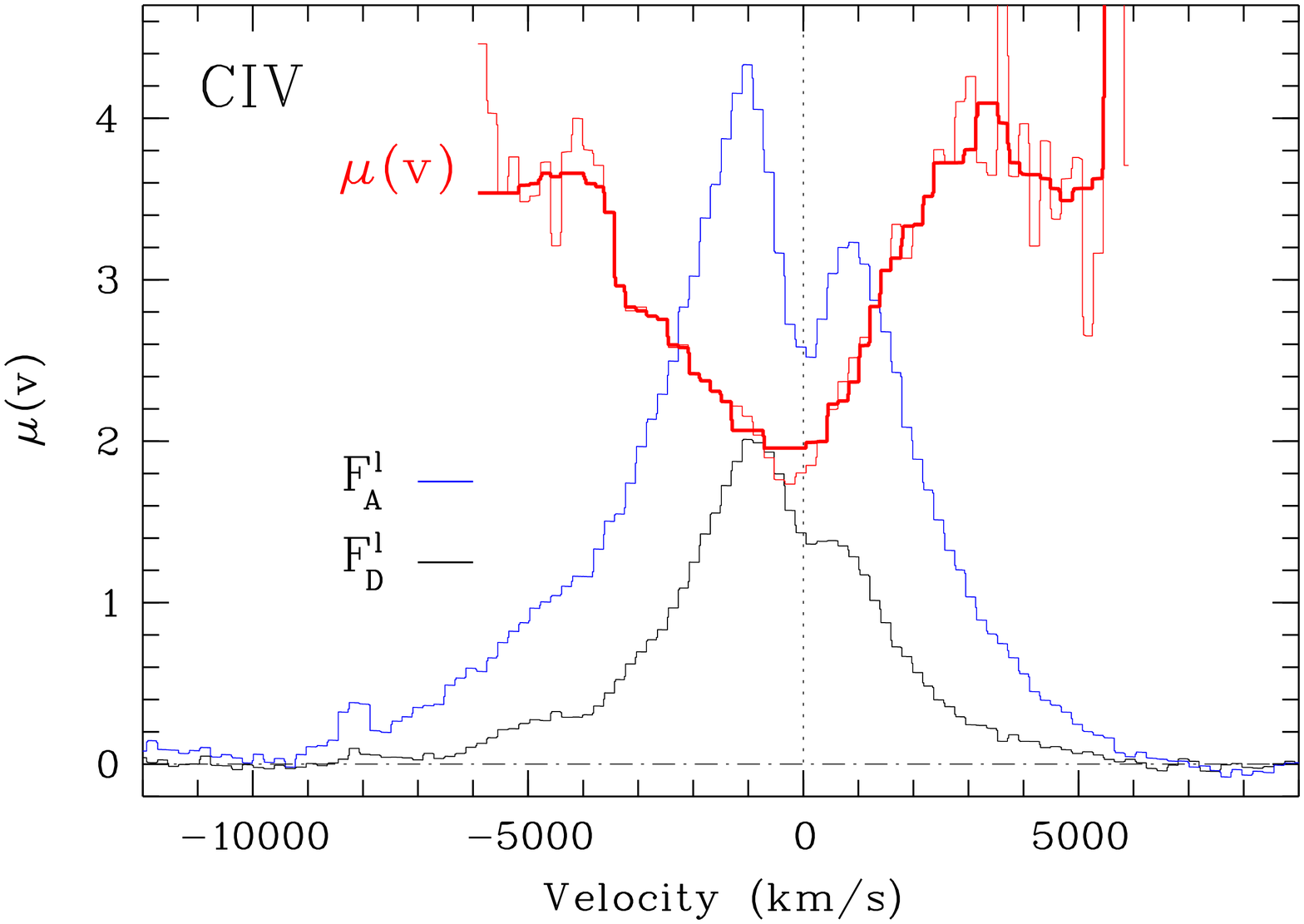}\includegraphics*[trim={30 0 0 0 mm},clip]{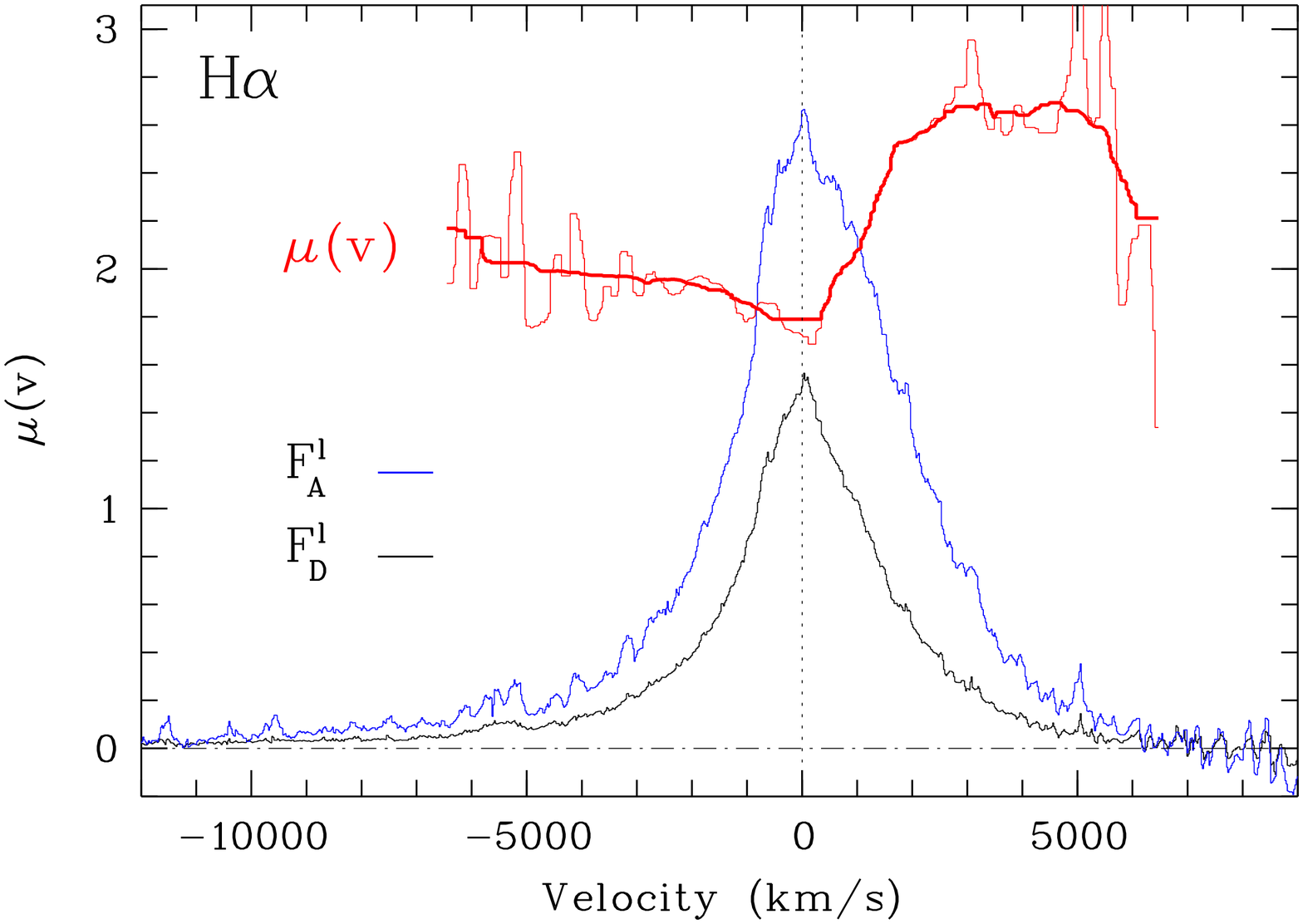}}
\caption{Flux density ratio  $\mu(v) = F^l_{\text A} / (M \times F^l_{\text D}),$ where $M = 1$, is illustrated  as a function of the Doppler velocity for the \ion{C}{iv} $\lambda$1549\AA\ and H$\alpha$ emission lines observed in the lensed quasar Q2237+0305 (red lines). $\mu(v)$ is shown after smoothing the line profiles with a ${220}$~km~s$^{-1}$ wide median filter and a ${1500}$~km~s$^{-1}$ wide median filter (thin and thick red lines, respectively). Outside the displayed velocity range, $\mu(v)$ becomes unreliable due to the low line fluxes. The continuum-subtracted spectra of images A and D ($F^l_{\text A}$ and $F^l_{\text D}$, respectively) are superimposed for comparison (blue and black lines, in arbitrary units).}
\label{fig:muwave_q2237}
\end{figure*}

Quasi-simultaneous near-infrared and visible spectra of the four components of Q2237+0305 have been obtained in October 2005 \citep{2008Eigenbroda,2016Braibant}. Observations, data reduction, and a first analysis focusing on microlensing are described in detail in \citet{2016Braibant}. These spectra show that image A of Q2237+0305 is affected by a strong microlensing magnification effect that clearly distorts the \ion{C}{iv} $\lambda$1549\AA\ and H$\alpha$ line profiles. Long-term monitoring of the system between 2000 and 2008 \citep[e.g.,][]{2015Mediavillab} shows that image D is most likely free of microlensing-induced variability. We then consider the spectrum of image D as the non-microlensed reference spectrum. The A/D macro-amplification ratio is $M =1.0 \pm 0.1$ \citep{2000Agol}. The microlensing-induced magnification factors of the continuum were estimated at the wavelengths of the \ion{C}{iv} and the H$\alpha$ emission lines in \citet{2016Braibant}, after correcting image D for the differential extinction. These factors, denoted $\mu^{cont}$, are reported in Table~\ref{tab:indices}.

\begin{table}[]
\caption{Measured magnification and distortion indices}
\label{tab:indices}
\centering
\begin{tabular}{lcccc}
\hline\hline
  Line  & $\mu^{cont}$ & $\mu^{BLR}$ & $WCI$ & $RBI$ \\
\hline
\ion{C}{iv}         & 3.05$\pm$0.25 & 2.51$\pm$0.27 & 1.57$\pm$0.05 & 0.05$\pm$0.03 \\
H$\alpha$   & 2.45$\pm$0.25 & 2.22$\pm$0.13 & 1.48$\pm$0.06 & 0.11$\pm$0.03 \\
\hline
\end{tabular}
\end{table}

\begin{figure*}
\hspace*{1.2cm} \mbox{{\large \ion{C}{iv} \ $q$=3} \hspace*{7.3cm} {\large \ion{C}{iv} \ $q$=1.5}}
\centering
\resizebox{\hsize}{!}{%
\includegraphics*[trim={0 0  0 0mm},clip]{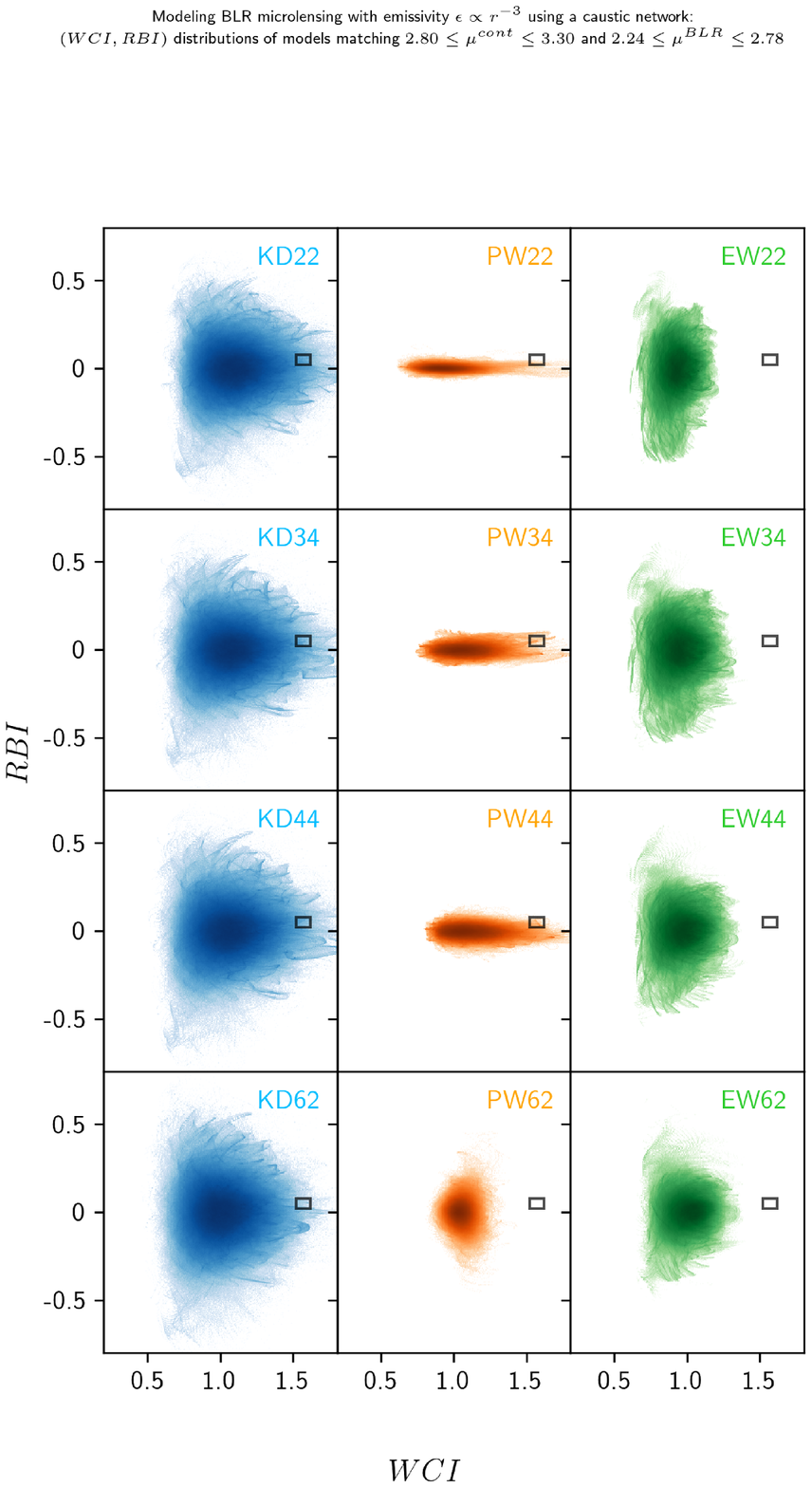}%
\includegraphics*[trim={12 0 0 0mm},clip]{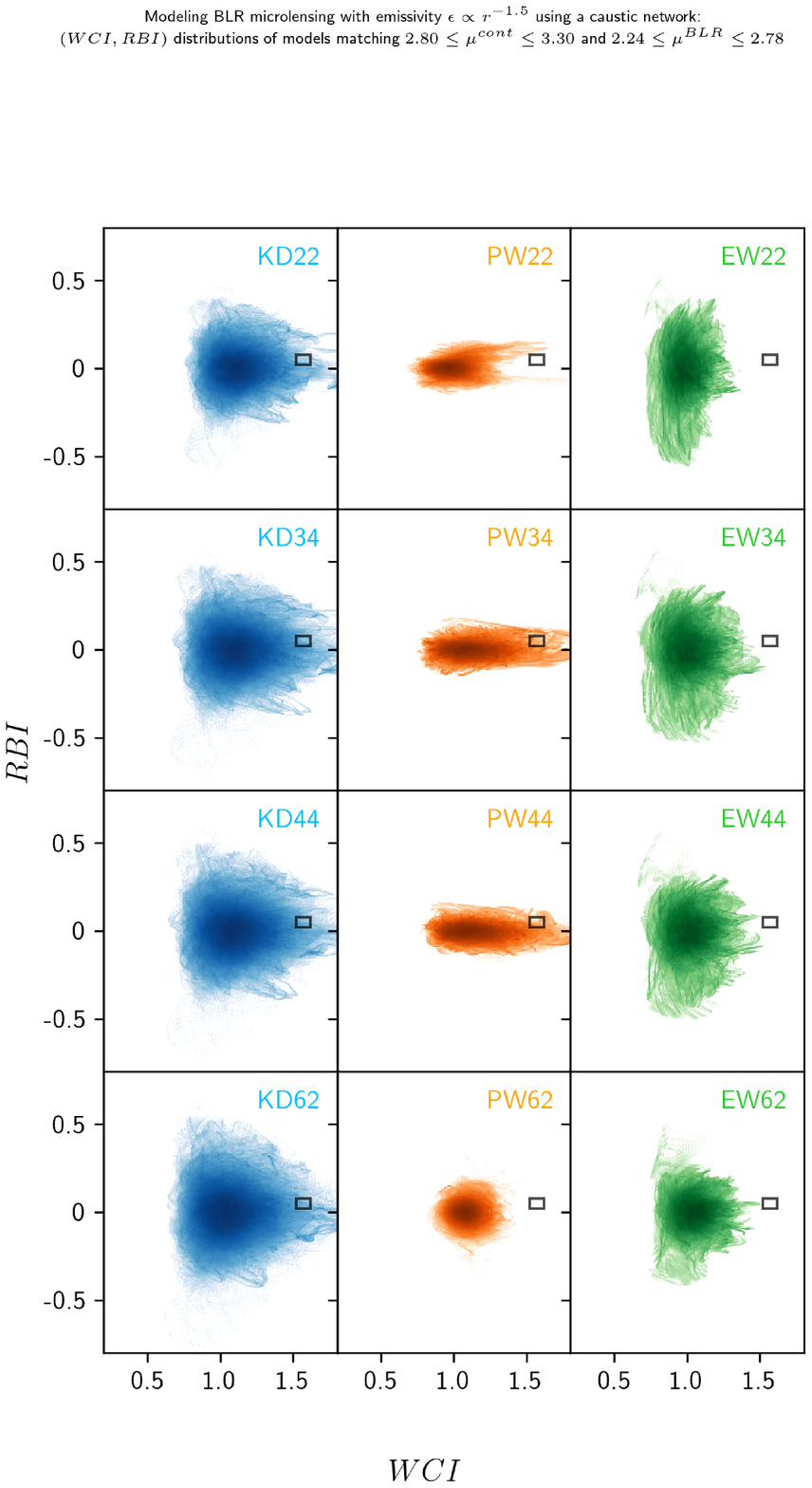}}
\caption{Two-dimensional histograms of simulated $WCI,RBI$ for the \ion{C}{iv} emission line. These indices were measured from simulated line profiles that arise from the BLR models KD, PW, and EW seen at inclinations 22\degr, 34\degr, 44\degr, and 62\degr. The BLR models with an emissivity $\epsilon_0 \, (r_{\text{in}}/r)^q$ that sharply decreases with radius, i.e., $q=3$, are illustrated in the left panel, while those that are characterized by a more slowly decreasing emissivity, i.e., $q=1.5$, are illustrated in the right panel. Each model also contains a continuum-emitting disk seen under the same inclination. All continuum source radii and BLR sizes used in the simulations are considered, but only the models that reproduce the $\mu^{cont}$ and $\mu^{BLR}$ values measured for the  \ion{C}{iv} emission line are shown, i.e.,  $2.8 \leq \mu^{cont} \leq 3.3$ and $2.24 \leq \mu^{BLR} \leq 2.78$ (Table~\ref{tab:indices}). The color map is logarithmic. The $WCI$ and $RBI$ values measured for \ion{C}{iv} are plotted with their uncertainties as a small rectangle superimposed on the simulated $WCI,RBI$ distributions.}
\label{fig:wcirbi_civ}
\end{figure*}

\begin{figure*}
\hspace*{1.3cm} \mbox{{\large H$\alpha$ \ $q$=3} \hspace*{7.4cm} {\large H$\alpha$ \ $q$=1.5}}
\centering
\resizebox{\hsize}{!}{%
\includegraphics*[trim={0 0 0 0 mm},clip]{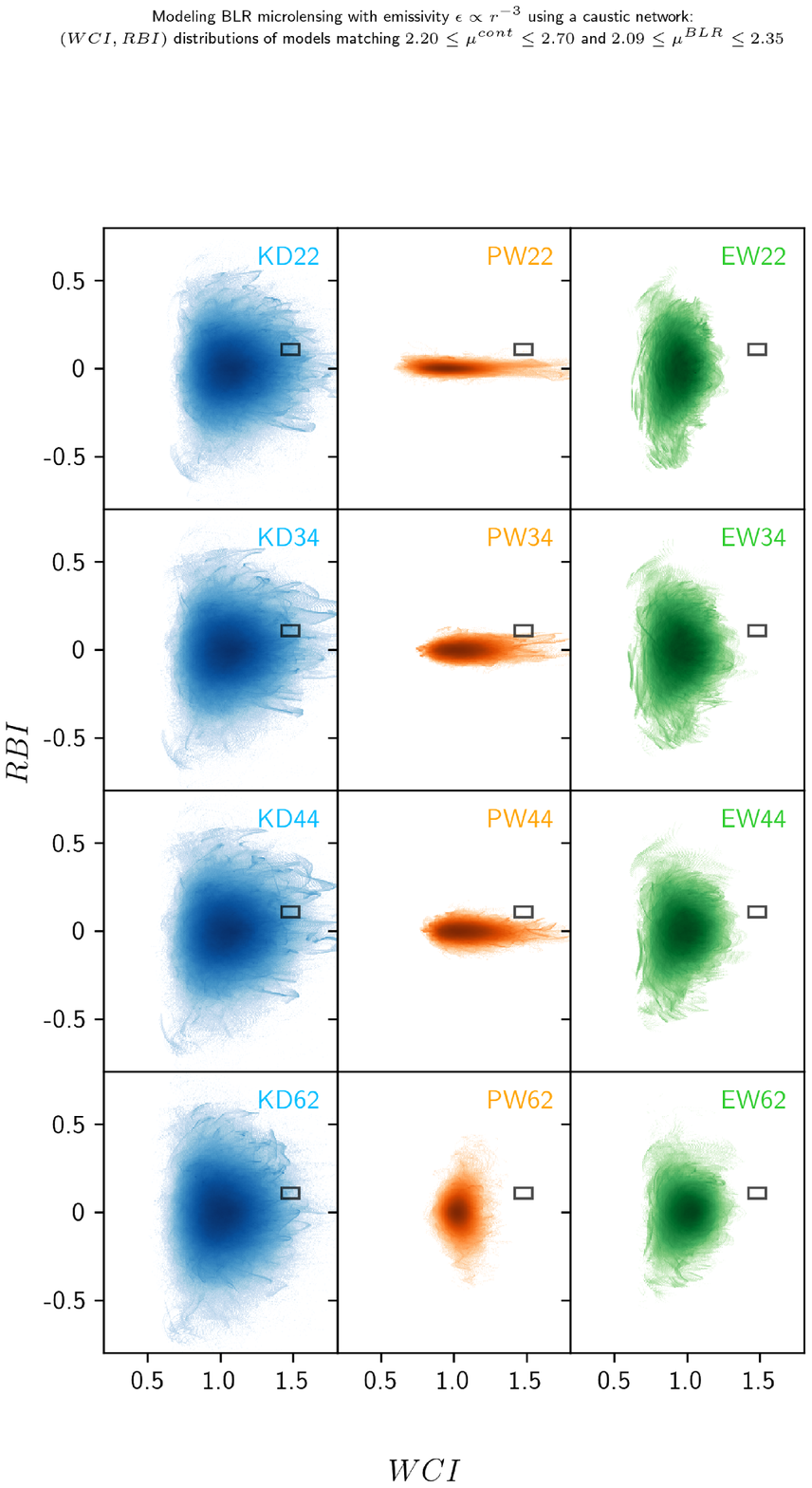}%
\includegraphics*[trim={12 0 0 0 mm},clip]{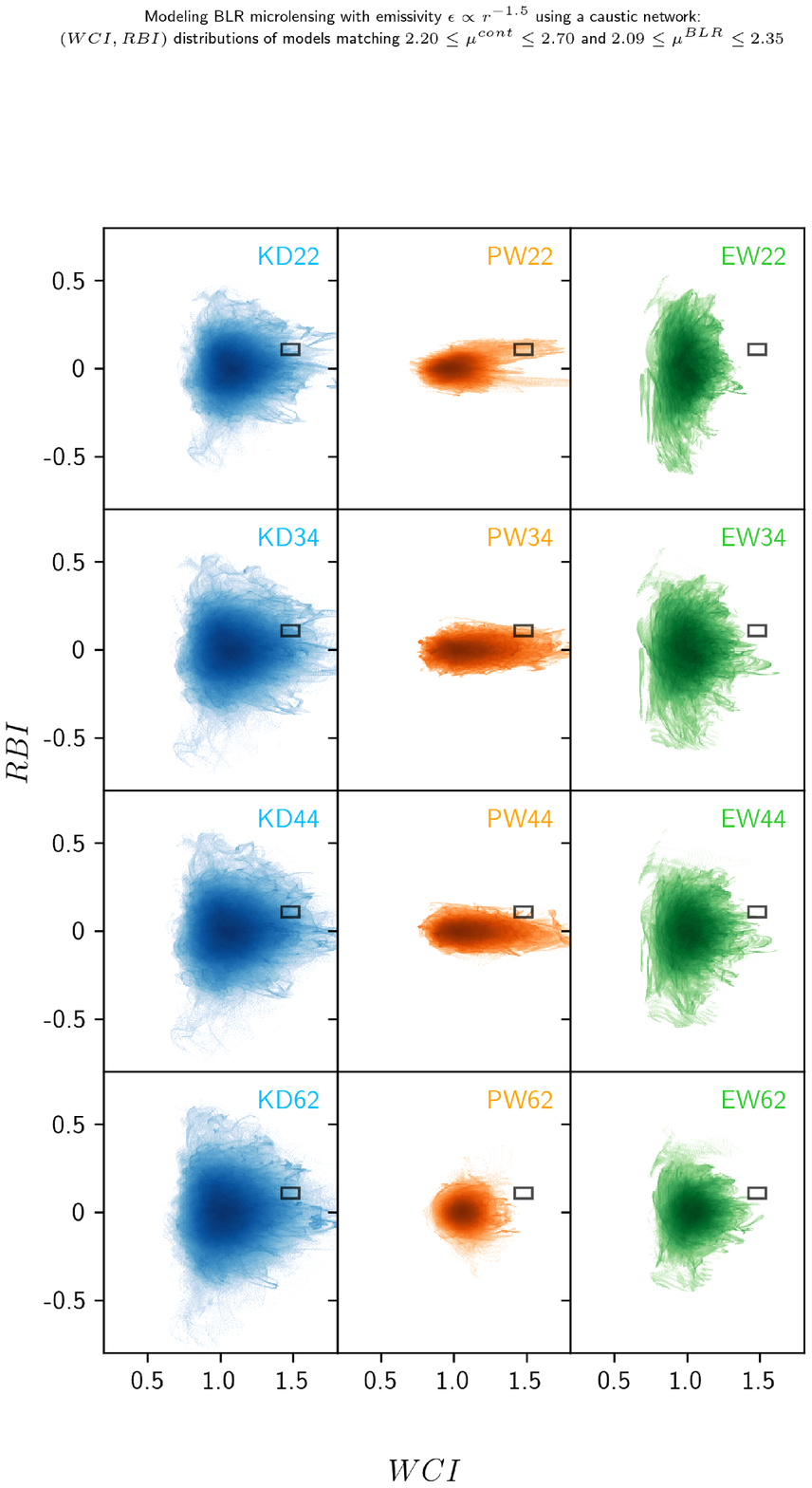}}
\caption{Two-dimensional histograms of simulated $WCI,RBI$ for the H$\alpha$ emission line. Same as Fig.~\ref{fig:wcirbi_civ}, but only the models that reproduce the $\mu^{cont}$ and $\mu^{BLR}$ values measured for the H$\alpha$ emission line are shown, i.e.,  $2.2 \leq \mu^{cont} \leq 2.7$ and $2.09 \leq \mu^{BLR} \leq 2.35$ (Table~\ref{tab:indices}). The $WCI$ and $RBI$ values measured for H$\alpha$  are plotted with their uncertainties as a small rectangle superimposed on the simulated $WCI,RBI$ distributions.}
\label{fig:wcirbi_ha}
\end{figure*}

The $\mu ^{BLR}$, $RBI$, and $WCI$ indices that quantify the effect of microlensing on the broad emission line profiles were estimated following \citet{2017Braibant} and \citet{2019Hutsemekers}. The total microlensing-induced magnification of an emission line, $\mu^{BLR}$, is obtained by integrating over the line profile the ratio between the emission line flux densities in images A and D,  $F^l_{\text{A}}$ and $F^l_{\text{D}}$ respectively, after subtraction of the local continuum and correction for the A/D macro-amplification ratio, that is,
\begin{equation}
\mu^{BLR} = \frac{1}{M} \frac{ \int_{v_{-}}^{v_{+}} \, F^l_{\text{A}} \, (v) \, dv}{\int_{v_{-}}^{v_{+}} F^l_{\text{D}} \, (v) \, dv}  \;  .
\label{eq:cmp-mublr}
\end{equation}
The red/blue $RBI$ and wings/core $WCI$ indices characterizing the line profile distortions are estimated using
\begin{equation}
RBI = \frac{\int_0^{v_{+}} \log \left( \mu(v) \right) dv}{\int_0^{v_{+}} dv} - \frac{\int_{v_{-}}^0 \log \left( \mu(v) \right) dv}{\int_{v_{-}}^0 dv} \; 
\label{eq:bluered_param}
\end{equation}
and
\begin{equation}
WCI = \frac{\int_{v_{-}}^{v_{+}} \mu(v)/ \mu(v=0) \, \, dv}{\int_{v_{-}}^{v_{+}} dv} \; 
\label{eq:wingcore_param}
\end{equation}
where
\begin{equation}
\mu \, (v) =  \frac{1}{M}  \frac{F^l_{\text{A}} \, (v) }{F^l_{\text{D}} \, (v)} \; .
\label{eq:muv}
\end{equation}
$RBI$ is sensitive to the asymmetry of the line profile deformations. It takes non-null values when the effect of microlensing on the blue and red parts of the line is asymmetric. $WCI$ indicates whether the whole emission line is on average more or less affected by microlensing than its center.  As they are defined, both $RBI$ and $WCI$ are independent of $M$. 

The magnification profile $\mu(v)$ is plotted in Fig.~\ref{fig:muwave_q2237} for the \ion{C}{iv} and the H$\alpha$ emission lines. Where the line flux is too low, $\mu(v)$ becomes unreliable. The indices are therefore computed over a restricted velocity range ([$-$5900,+5900] km s$^{-1}$ for \ion{C}{iv} and [$-$6450,+6450] km s$^{-1}$ for H$\alpha$) that nevertheless encompasses most of the line profiles. The $\mu ^{BLR}$, $WCI$, and $RBI$ indices are reported in Table~\ref{tab:indices}. The errors on the indices are obtained by propagating the uncertainty of the line flux densities. This uncertainty is computed as the quadratic sum of the error on the total (line + continuum) flux density and the error on the continuum estimate, the latter being taken as the standard deviation of the continuum flux on each side of the emission lines. The exact value of the indices depends on several parameters such as the adopted underlying continuum,  the velocity range, and the systemic redshift that defines the line center. The measured values are nevertheless stable within the quoted uncertainties for small variations of these parameters. They are also similar when using different smoothing values like those illustrated in Fig.~\ref{fig:muwave_q2237}.

It is immediately clear from the shape of $\mu(v)$ and the values of the $WCI$ and $RBI$ indices that the \ion{C}{iv} and the H$\alpha$ emission lines are differently affected by the microlensing effect, as already noted in \citet{2016Braibant}. The significantly positive value of $RBI$ for H$\alpha$ indicates that microlensing magnifies the receding velocities to a larger extent than the approaching ones. $WCI$ larger than one in both lines indicates a significantly higher magnification of the line wings with respect to the line center.

\section{Comparison with simulations}
\label{sec:modeling}

\subsection{Microlensing simulations}

We  investigated the effect of gravitational microlensing on broad emission line profiles and the underlying continuum by convolving in the source plane the emission from representative BLR models with microlensing magnification maps. For the BLR models, we considered a rotating Keplerian disk (KD), as well as biconical polar (PW) and equatorial (EW) radially accelerated winds with inclinations with respect to the line of sight $i$ = 22\degr, 34\degr, 44\degr, 62\degr.  We investigated $\text{nine}$ inner radius values for each BLR model: $r_{\text{in}}=$ 0.1, 0.125, 0.15, 0.175, 0.2, 0.25, 0.35, 0.5, and 0.75 $r_E$, where $r_E$ is the microlensing Einstein radius. To compute the Einstein radius, we adopted a flat $\Lambda$CDM cosmology with $H_0 = 68$ km~s$^{-1}$ Mpc$^{-1}$ and $\Omega_m$ = 0.31. The source and lens redshifts are $z_S$ = 1.695 and $z_L$ = 0.0394, respectively \citep{1985Huchra}, so that the Einstein radius is $r_E = 3.26 \times 10^{-2}$ pc in the source plane for an average microlens mass of 0.3 M$_{\odot}$. The outer radius was fixed to $r_{\text{out}} = 10 \, r_{\text{in}}$. This value largely encompasses the mean and half-light radii of the BLR models (Appendix A). The BLR models are assumed to have an emissivity $\epsilon = \epsilon_0 \, (r_{\text{in}}/r)^q$ that either sharply decreases with radius, that is, $q=3$, or more slowly, that is, $q=1.5$. Twenty BLR monochromatic images corresponding to twenty spectral bins in the line profile were produced using the radiative transfer code STOKES \citep{2007Goosmann,2012Marin,2014Goosmann}. Each model also contains a continuum-emitting uniform disk seen under the same inclination as the BLR, and with an outer radius fixed at $\text{nine}$ different values: $r_s =$ 0.1, 0.15, 0.2, 0.25, 0.3, 0.4, 0.5, 0.6, and 0.7 $r_E$. These values sample the range of continuum half-light radii $r_{1/2}$ measured for Q2237+0305 and collected by \citet{2011Sluse}, using $r_s \simeq 1.4 \,r_{1/2}$ for a uniform disk. Only models with $r_{\text{in}} \geq r_s$ were considered.  For a full description of the models and additional details, we refer to \citet{2017Braibant}.

The effect of microlensing on the BLR was modeled using a caustic network specific to image A of Q2237+0305.  The magnification map was computed using the \texttt{microlens} ray-tracing code \citep{1999Wambsganss}, considering a convergence $\kappa_s = $ 0.394  for matter in compact objects and $\kappa_c = $ 0 for continuously distributed matter, with an external shear $\gamma =$ 0.395 \citep{2004Kochanek}. The map extends over a $200 \, r_E \times 200 \, r_E$ area of the source plane and is sampled by $20000 \times 20000$ pixels. To limit the impact of preferential alignment between the symmetry axes of the BLR models and the caustic network, the map was rotated by 0\degr, 30\degr, 45\degr, 60\degr\ , and 90\degr , and only the central $10000 \times 10000$ pixel parts of the rotated maps were finally used.

Distorted line profiles were obtained from the convolution of the monochromatic images of the BLR by the magnification maps, and were computed for each position of the BLR on the magnification maps, generating $\sim 10^8$ simulated profiles per map and BLR model. Given the huge amount of simulated line profiles, we focused our analysis on the indices $\mu^{cont}$, $\mu^{BLR}$, $RBI$, and $WCI$ that characterize the continuum and line profile magnification and distortions, and which can be directly compared with the observations.

\subsection{$WCI,RBI$ distributions}

Figs.~\ref{fig:wcirbi_civ} and~\ref{fig:wcirbi_ha} display for \ion{C}{iv} and H$\alpha$  the $WCI,RBI$ distributions of the simulated microlensed line profiles that arise from the magnification of the different BLR models by the caustic network. Only simulations that match the $\mu^{cont}$ and $\mu^{BLR}$ values are represented. The BLR models are grouped according to their geometry/kinematics, inclination, and emissivity. The ranges of $WCI$ and $RBI$ values that correspond to the observed distortion of the line profiles (Table~\ref{tab:indices}) are indicated by a small rectangle superimposed on the simulated $WCI,RBI$ distributions.

We immediately see that a wide range of $WCI,RBI$ values is generated. With the KD BLR model, high values of both $WCI$ and $RBI$ are generated. With the PW and EW models, either high $WCI$ or high $RBI$ values can be produced, respectively. Simulations performed with the $q=3$ and $q=1.5$ emissivity indices are very similar, as are the simulations performed for \ion{C}{iv} and H$\alpha$, which only differ by the constraints on $\mu^{cont}$ and $\mu^{BLR}$.

The KD model can reproduce the observed magnification and distortion indices regardless of the inclination and emissivity law. The EW model is rejected in all cases by the observations. The PW model can reproduce the \ion{C}{iv} line deformation for the inclinations 34\degr\ and 44\degr, and the H$\alpha$ line deformation for the inclinations  22\degr\ and 34\degr\ with the $q=1.5$ emissivity law. The comparison of the observed indices with the simulations thus provide clear constraints on the geometry and kinematics of the BLR in Q2237+0305. The constraints are stronger than those derived for the lensed quasar HE0435-1223 (see Fig.~2 of \citealt{2019Hutsemekers}) due to a stronger microlensing effect ($\mu^{BLR}$ is significantly higher in  Q2237+0305 than in 
HE0435-1223).

\subsection{Probability of the different models}

\begin{table}
\caption{Probability (in \%) of the different BLR models for H$\alpha$ and \ion{C}{iv}}
\label{tab:proba1}
\centering
\begin{tabular}{lcccc}
\hline\hline
   &  \multicolumn{4}{c}{H$\alpha$ } \\
\hline
          & KD & PW & EW & ALL \\
\hline
          22\degr         & 24 &  2 &  0 & 26  \\
          34\degr         & 27 &  7 &  0 & 34  \\
          44\degr         & 24 &  2 &  0 & 26  \\
          62\degr         & 13 &  0 &  0 & 13  \\
          All $i$         & 88 & 11 &  0 &     \\
\hline\hline
   &  \multicolumn{4}{c}{\ion{C}{iv}} \\
\hline
          & KD & PW & EW & ALL \\
\hline
          22\degr         & 22 &  1 &  0 & 23 \\
          34\degr         & 23 & 10 &  0 & 33 \\
          44\degr         & 20 & 15 &  0 & 35 \\
          62\degr         & 10 &  0 &  0 & 10 \\
          All $i$         & 75 & 26 &  0 &    \\
\hline
\end{tabular}
\end{table}

In order to identify the BLR models that best fit the observations, we computed the relative posterior probability that a given model $(G, i)$ (where $G$ = KD, PW, or EW, and $i$ = 22\degr, 34\degr, 44\degr, or 62\degr) can more easily reproduce the four observables $\mu^{cont}$, $\mu^{BLR}$, $RBI$, and $WCI$. The method is  described in detail in \citet{2019Hutsemekers}. Practically, we computed the likelihood of the observables for each set of parameters characterizing the simulations. We then marginalized the likelihood over $r_s$, $r_{\text{in}}$, and $q$, as well as over the microlensing parameters. Because the different BLR models share the same parameters and associated priors, we can quantify their relative efficiency to reproduce the data by comparing their likelihoods, that is, by normalizing the marginalized likelihood by the sum of the likelihoods associated with each model $G$ for each inclination $i$. This yields the relative probability of the different models, the values of which are given in Table~\ref{tab:proba1} for the H$\alpha$ and \ion{C}{iv} BLRs independently.

\begin{table}
\caption{Probability (in \%) of [H$\alpha$/\ion{C}{iv}] BLR model combinations}
\label{tab:proba2}
\centering
\begin{tabular}{lccccc}
\hline\hline
&  \multicolumn{5}{c}{All $r_s$}  \\
\hline
          & KD/KD & KD/PW & PW/KD & PW/PW & ALL \\
\hline
          22\degr         & 20 &  0 &  0 &   3  & 23 \\
          34\degr         & 22 &  8 &  1 &  10  & 41 \\
          44\degr         & 18 &  6 &  1 &   3  & 28 \\
          62\degr         &  7 &  0 &  0 &   0  &  7 \\
          All $i$         & 67 & 14 &  2 &  16  &    \\
\hline\hline
 &  \multicolumn{5}{c}{$r_s (\lambda_{H\alpha}) >  2\times r_s (\lambda_{\ion{C}{iv}})$} \\
\hline
          & KD/KD & KD/PW & PW/KD & PW/PW & ALL  \\
\hline
          22\degr         & 22 &  0 &  0 &   2  & 24 \\
          34\degr         & 22 &  7 &  0 &   9  & 38 \\
          44\degr         & 22 &  7 &  0 &   2  & 31 \\
          62\degr         &  8 &  0 &  0 &   0  &  8 \\
          All $i$         & 74 & 14 &  0 &  13  &  \\
\hline
\end{tabular}
\end{table}

\begin{figure}[t]
\centering
\resizebox{\hsize}{!}{%
  \includegraphics*{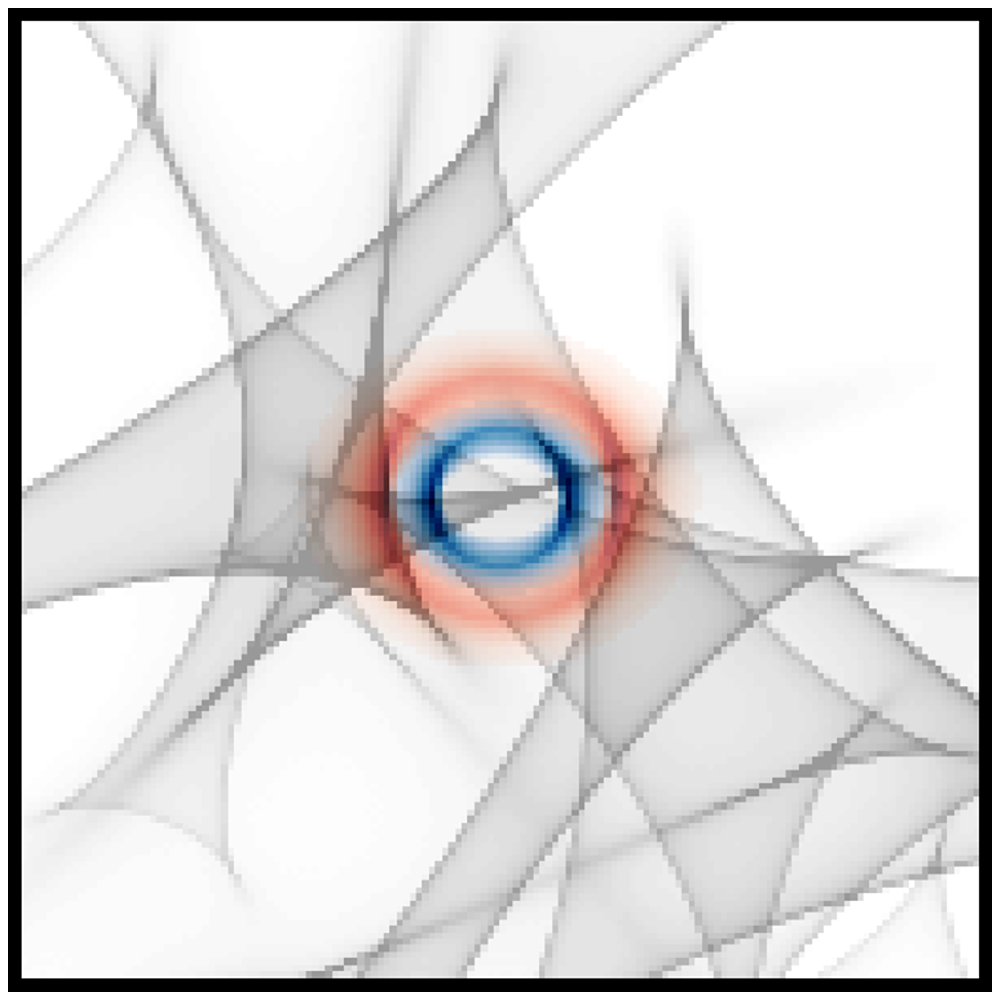}$ $
  \includegraphics*{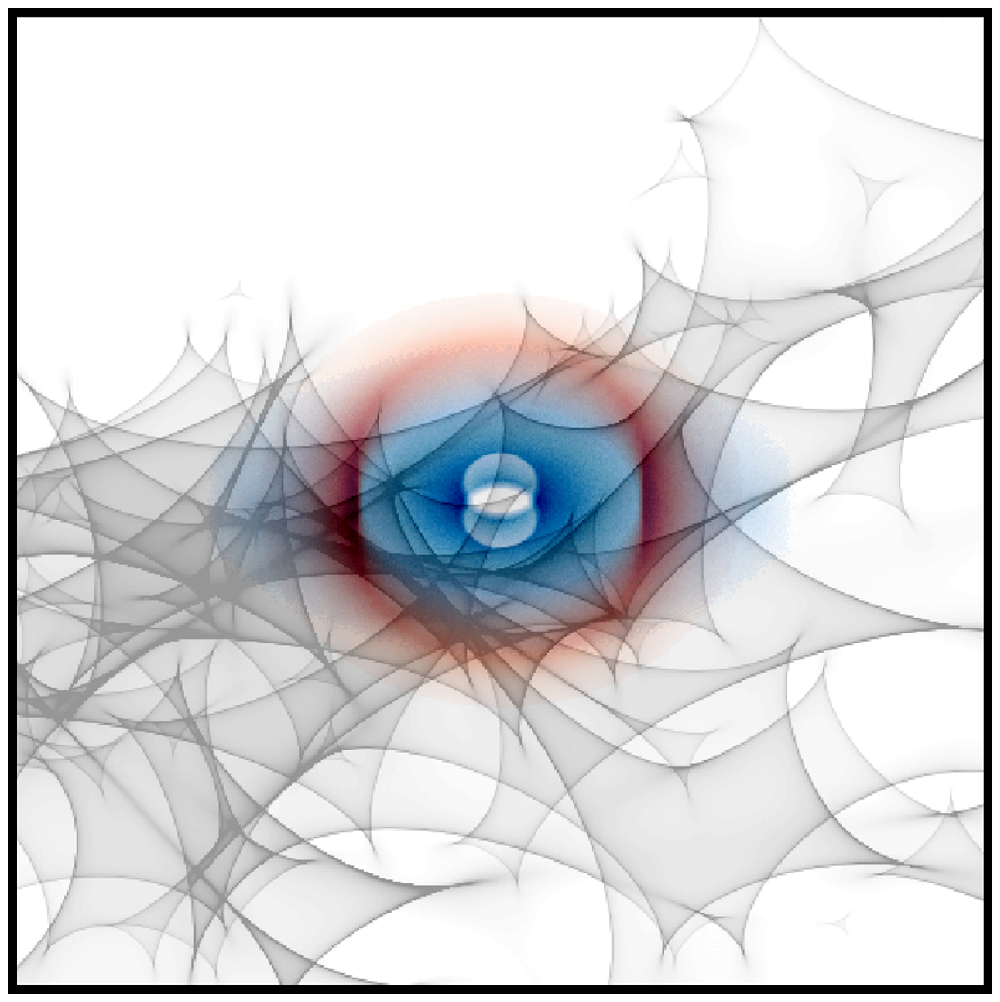}}%
\caption{H$\alpha$ (in red) and \ion{C}{iv} (in blue) BLR intensity maps superimposed on the magnification map with 0\degr\ rotation. These combinations of models and maps reproduce the measurements within the uncertainties. {\bf Left:} The H$\alpha$ BLR model is characterized by the following parameters: KD, $i = 34\degr$, $q = 3$, $r_s$ = 0.15 $r_E$ , and $r_{\text{in}}=$ 0.175 $r_E$, and the \ion{C}{iv} BLR model is characterized by KD, $i = 34\degr$, $q = 3$, $r_s$ = 0.10 $r_E$ , and $r_{\text{in}}=$ 0.10 $r_E$. The size of the map is 1.5 $r_E$ $\times$ 1.5 $r_E$. {\bf Right:} The H$\alpha$ BLR model is characterized by KD, $i = 34\degr$, $q = 3$, $r_s$ = 0.15 $r_E$ , and $r_{\text{in}}=$ 0.75 $r_E$, and the \ion{C}{iv} BLR model is characterized by  PW, $i = 34\degr$, $q = 1.5$, $r_s$ = 0.10 $r_E$ , and $r_{\text{in}}=$ 0.25 $r_E$. The size of the map is 5~$r_E$ $\times$ 5~$r_E$.}
\label{fig:blroncaustic}
\end{figure}

We see that EW BLR models are totally rejected. The KD models are the most likely for both the H$\alpha$ and \ion{C}{iv} lines. PW models are also possible, especially for the high-ionization \ion{C}{iv} BLR. In all cases, the most likely inclinations are 34\degr\ and 44\degr. Because the exact value of $RBI$ for the \ion{C}{iv} line could be affected by a global blueshift of the line, we also computed the model probabilities using $RBI$ = 0, that is, a symmetric deformation of the \ion{C}{iv} line. We find probabilities of 66\% and 34\% for the KD and PW (all $i$) models, respectively. These values do not strongly differ from the values given in Table~\ref{tab:proba1} (75\% and 26\%, respectively), thus indicating that these results are robust.

Because the H$\alpha$ and \ion{C}{iv} BLRs are expected to share the same center and inclination, we also computed the probability of combinations of H$\alpha$ and \ion{C}{iv} BLR models that have the same position on the caustic map as well as identical inclinations. The EW models being rejected, we only considered the combinations of KD and PW models. Results are given in Table~\ref{tab:proba2}. The model combinations with KD for both the H$\alpha$ and \ion{C}{iv} BLRs are clearly favored. Other combinations involving PW for \ion{C}{iv} are also possible. The most likely inclinations are 34\degr\ and 44\degr. We also added an independent constraint on the relative size of the continuum source at the wavelengths of H$\alpha$ and \ion{C}{iv} derived from microlensing-induced time variations. This constraint affects the BLR models through the condition  $r_{\text{in}} \geq r_s$. Between 1500~\AA\ and 3000~\AA\ restframe, the size of the continuum source was found to scale as $\lambda^{p}$ , where $p$ is between 0.8 and 1.3 \citep{2008Eigenbrodb,2016Munoz,2020Goicoechea}. We could thus expect a factor 4 between the size of the continuum source at the wavelength of H$\alpha$ and its size at the wavelength of \ion{C}{iv}. However, because no measurement is available at the wavelength of H$\alpha$, we conservatively used $r_s (\lambda_{H\alpha}) >  2\times r_s (\lambda_{\ion{C}{iv}})$. As seen in Table~\ref{tab:proba2}, this additional constraint does not change the results.

In Fig.~\ref{fig:blroncaustic} we show two representative examples of model and map combinations that within the uncertainties reproduce the observables $\mu^{cont}$, $\mu^{BLR}$, $RBI$, and $WCI$ measured for both H$\alpha$ and \ion{C}{iv}. In the left panel, the combination KD(H$\alpha$)/KD(\ion{C}{iv}) with $i=34\degr$ is illustrated. Due to the dense caustic network, a small difference in BLR size is sufficient to have a roughly symmetric magnification for the \ion{C}{iv} BLR and an asymmetric one for the H$\alpha$ BLR, considering the same BLR model. The right panel shows a KD(H$\alpha$)/PW(\ion{C}{iv}) combination at $i=34\degr$. In this case, the BLR size is larger and more caustics are involved in a complex magnification pattern of the BLR. Multi-epoch data should allow us to disentangle these combinations. In these examples the H$\alpha$ BLR appears larger than the \ion{C}{iv} BLR, but it should be emphasized that we could not derive any significant information on their relative sizes on the basis of our data set (Sect.~\ref{sec:size}).

\subsection{Size of the BLR and black hole mass}
\label{sec:size}

\begin{figure}[t]
\centering
\resizebox{\hsize}{!}{\includegraphics*{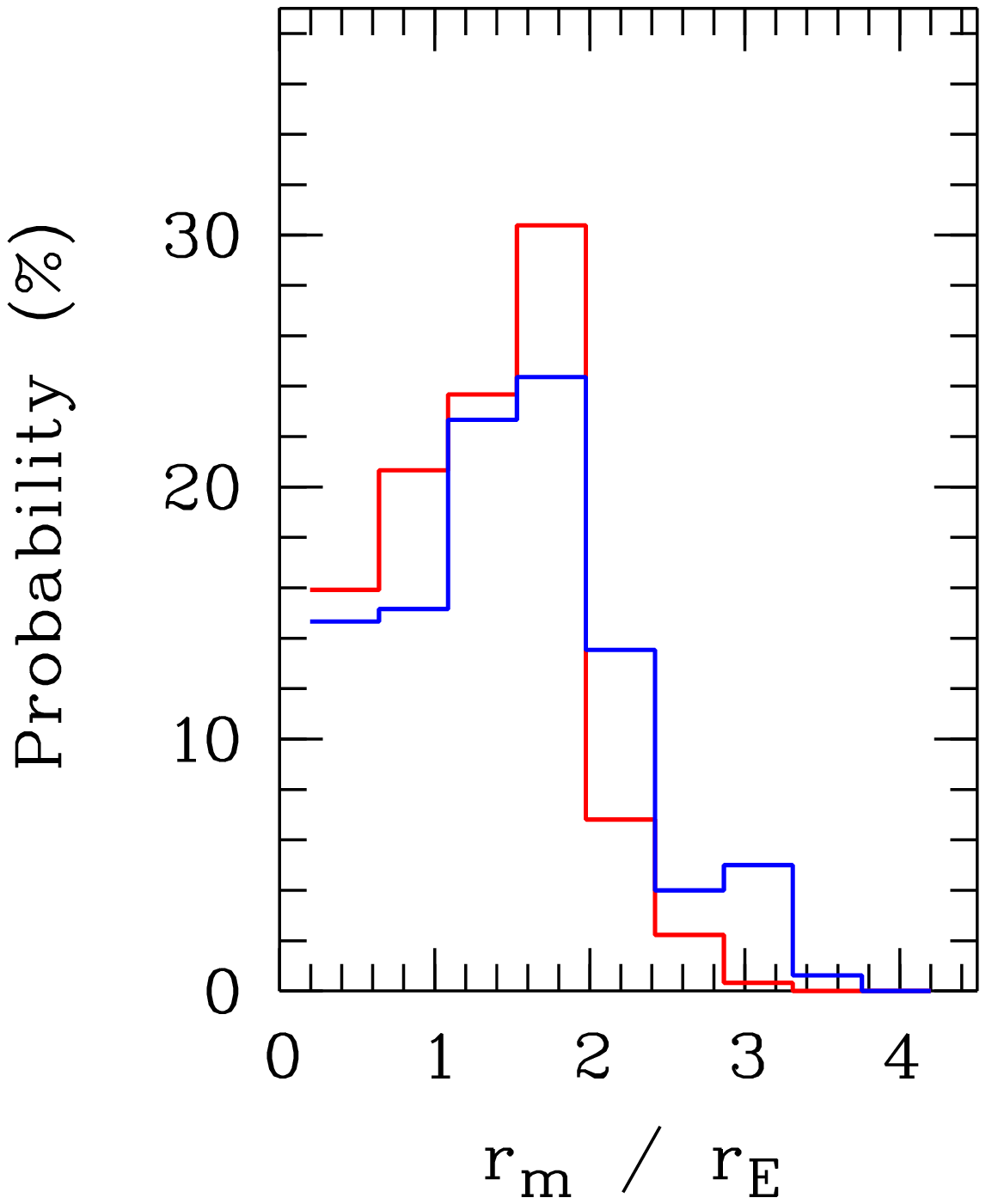}\includegraphics*{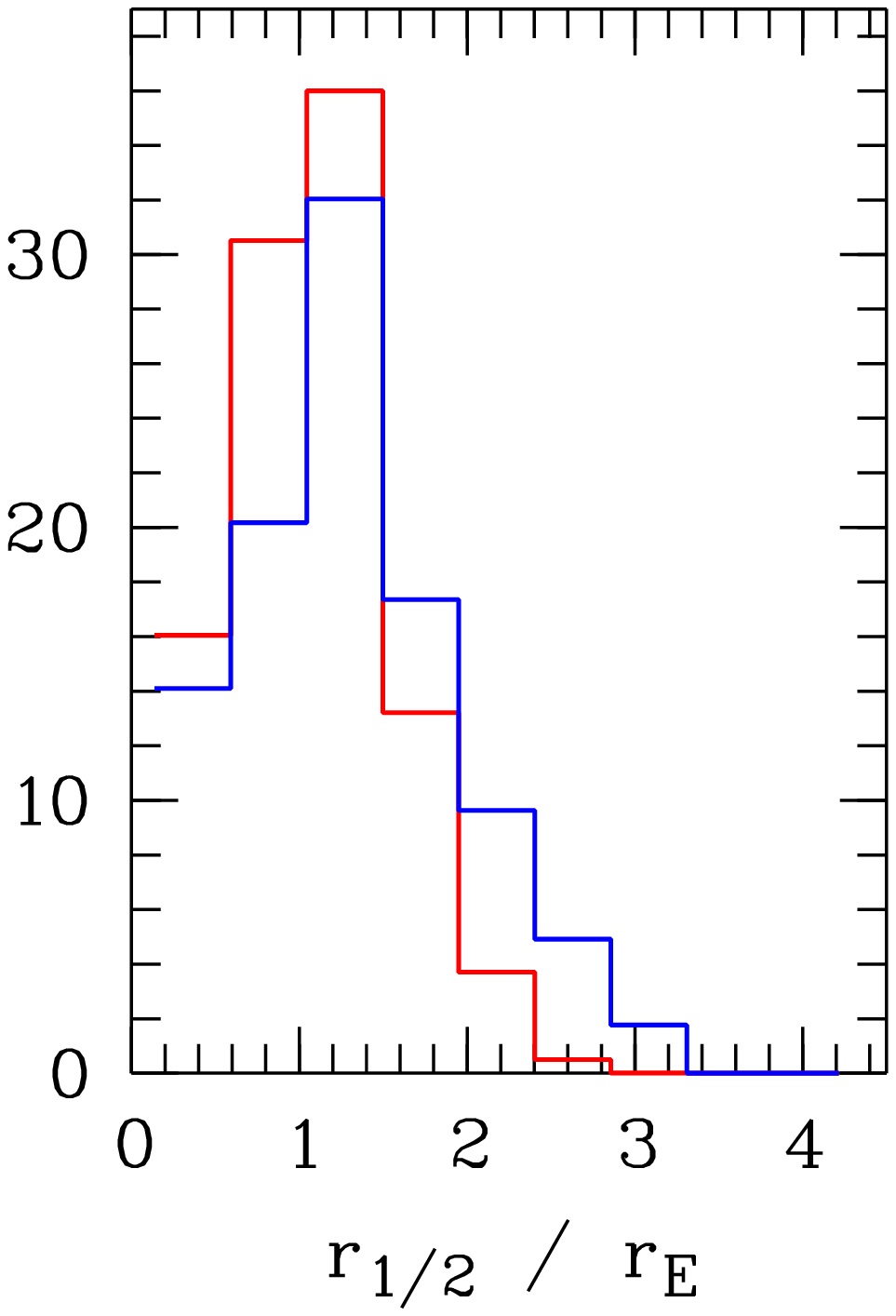}}
\caption{Relative probabilities (in percent) of the BLR mean radius $r_{\rm m}$ and half-light radius $r_{\rm 1/2}$ in $r_E$ units. Blue refers to the \ion{C}{iv} BLR, and red to the H$\alpha$ BLR.}
\label{fig:rblrmehl}
\end{figure}

By marginalizing over all parameters but $r_{\text{in}}$, we can estimate the most likely BLR radius. However, $r_{\text{in}}$ does not properly represent the size of the BLR, which also depends on the light distribution. We therefore computed the mean and half-light radii for the different models as described in Appendix~\ref{sec:appendix}, and estimated their relative probabilities in nine bins between the minimum and maximum values. The resulting probabilities are shown in Fig.~\ref{fig:rblrmehl}. There is no significant difference between the sizes of the \ion{C}{iv} and H$\alpha$ BLRs. From the probability distributions, we estimate the mean radius of the BLR to be $r_{\rm m} \simeq$ 1.4$\pm$0.6 $r_E$ $\simeq$ 54$\pm$23 light-days, and the half-light radius to be $r_{\rm 1/2} \simeq$ 1.2$\pm$0.5 $r_E$ $\simeq$ 47$\pm$19 light-days. The latter value is in excellent agreement with the \ion{C}{iv} BLR half-light radius derived from microlensing-induced time variability of the \ion{C}{iv} line \citep{2011Sluse}.

Our estimate of the BLR radius may be compared to the radius-luminosity ($R-L$) relations derived from reverberation mapping. Using the relations given by \citet{2021Kaspi} for the \ion{C}{iv} BLR, and the luminosity reported by \citet{2011Assef} (corrected for our cosmological parameters, and assuming $\lambda L_{\lambda}$(1350\AA ) = $\lambda L_{\lambda}$(1450\AA ) as in \citealt{2006Vestergaard}), we obtain $R$(\ion{C}{iv}) $\simeq$ 55$\pm$30 light-days, which is in excellent agreement with our value. As shown by \citet{2017Grierb}, the $R-L$ relation for the Balmer line BLR is more complex: many measurements lie below the canonical relation of \citet{2013Bentz}, thus possibly indicating selection effects or different BLR sizes in high-luminosity quasars. We translate our H$\alpha$ BLR size measurement into an H$\beta$ BLR size using the conversion $r_{\rm 1/2}$(H$\beta$) $\sim$ 0.7 $\times \, r_{\rm 1/2}$(H$\alpha$) $\sim$ 33 light-days, which is derived visually from Fig.~10 of \citet{2017Grierb}. This value is one order of magnitude lower than the size derived from the $R-L$ relation of \citet{2013Bentz}, which yields $R$(H$\beta$) $\sim$ 400 light-days using $\log [\lambda L_{\lambda}$(5100\AA )] $\simeq$ 46 \citep{2011Assef}. The microlensing estimate of the H$\beta$ BLR radius thus lies much below the canonical relation, but it does agree with the deviation reported by \citet{2017Grierb} for high-luminosity objects. Because our measurements are not affected by the same selection effects as reverberation mapping, this might point to a real physical effect, possibly related to the accretion rate \citep{2016Dub}. Indeed, Q2237+0305 has a high dimensionless accretion rate $\dot{\mathscr{M}}$ = 370, computed using Eq.~3 of \citet{2016Dub}.

Assuming virial motion, the mass of the black hole can be estimated using
\begin{equation}
\label{eq:mbh}
M_{BH} = f \, \frac{R_{BLR} \, V^2}{G}
,\end{equation}
where $V$ is the velocity full width at half maximum (FWHM) of the broad emission line, $f$ the virial factor and $G$ the gravitational constant. For a thin Keplerian disk (our most likely model), \citet{2008Decarli} give
\begin{equation}
f = (4 \, \sin^2i)^{-1} \, .
\end{equation}
For the broad lines observed in the spectra of image D (unaffected by microlensing), we measure FWHM(\ion{C}{iv}) = 3800$\pm$400 km~s$^{-1}$ and FWHM(H$\alpha$) = 4000$\pm$400  km~s$^{-1}$ by fitting a set of  broad and narrow lines. For the narrow lines, we use FWHM $\simeq$ 1070 km~s$^{-1}$ estimated from the [OIII] lines \citep{2010Greene}. The derived FWHMs of the \ion{C}{iv} and H$\alpha$ broad lines agree with those obtained by \citet{2011Assef} and \citet{2010Greene}. According to Eq.~\ref{eq:mbh}, similar FWHMs of the \ion{C}{iv} and H$\alpha$ broad lines are expected if the BLRs have similar radii, as observed. Using $i$ = 40$\pm$5\degr, $R_{BLR}$ = 47$\pm$19 light-days, and $V$ = 3900$\pm$500 km~s$^{-1}$, we finally obtain $M_{BH} = 9\pm5 \times 10^7$ M$_{\odot}$.

\section{Conclusions}
\label{sec:conclu}

We have compared quasi-simultaneous observations of the microlensing effect on the H$\alpha$ and \ion{C}{iv} broad emission lines in the quasar Q2237+0305 with simulations based on simple but representative BLR models and magnification maps. Our study shows that
\begin{itemize}
\item the observed microlensing effect, characterized by a set of four indices, can be reproduced with some of the considered BLR models,
\item the equatorial wind model is rejected,
\item for both the H$\alpha$ and the \ion{C}{iv} BLRs the most likely model is the Keplerian disk; models with a polar wind for the high-ionization \ion{C}{iv} BLR are possible, but less likely,
\item the most likely inclination of the system with respect to the line of sight is about 40\degr\ , as expected for a type~1 AGN and in agreement with the value derived from microlensing time series \citep{2010Poindexter},
\item the BLR half-light radius is $\simeq$ 47$\pm$19 light-days; the difference between the sizes the \ion{C}{iv} and H$\alpha$ BLRs cannot be estimated from our data set,
\item the \ion{C}{iv} BLR radius agrees with the value obtained from microlensing time series and with the $R-L$ relation derived from reverberation mapping, while the Balmer line BLR is one order of magnitude smaller than the value expected from the $R-L$ relation of \citet{2013Bentz}.
\end{itemize}

The different deformations of the H$\alpha$ and \ion{C}{iv} emission lines observed at the same epoch led \citet{2017Braibant} to suggest that the geometry and kinematics of the H$\alpha$ and \ion{C}{iv} BLRs were different. However, as illustrated in Fig.~\ref{fig:blroncaustic}, small differences in BLR sizes are sufficient to produce different microlensing signatures with the same BLR model when the caustic network is dense. The most likely model we found for the H$\alpha$ BLR in Q2237+305, the Keplerian disk, was also suggested in the lensed quasar HE0435-1223 using a similar analysis \citep{2019Hutsemekers}. This model does agree with the BLR models favored by reverberation mapping and near-infrared interferometry studies of other AGNs (Sect.~\ref{sec:intro}). Microlensing clearly allows us to independently constrain the BLR geometry and kinematics in lensed quasars. While interferometry is currently limited to a few nearby AGNs, and reverberation mapping to low-luminosity low-redshift objects due to the very long response time in luminous and high-redshift objects, microlensing can explore the BLR in bright distant quasars as far as they are lensed and show strong line deformations. As shown in Figs.~\ref{fig:wcirbi_civ} and~\ref{fig:wcirbi_ha}, high WCI values together with high RBI values can only be reproduced by the KD model. Single-epoch observations of strong, simultaneous, red/blue and wings/core distortions can thus be sufficient to discard the PW and EW models.  Multi-epoch data would nevertheless be needed to catch the most sensitive line profile deformations (or place limits on them) and to test more sophisticated BLR models, in particular, by making use of the full  $\mu(v)$ magnification profile.

\begin{acknowledgements}
We warmly thank Lorraine Braibant for her contribution to the earliest phases of this project. We also thank the referee for useful suggestions. This work was supported by the F.R.S.–FNRS under grants IISN~4.4503.19 and PDR~T.0116.21.
\end{acknowledgements}

\bibliographystyle{aa}
\bibliography{references}

\begin{appendix}
\section{BLR mean and half-light radii }
\label{sec:appendix}

The mean emission radius, $r_{\rm m}$, is defined as the radial distance weighted by the surface brightness of the BLR seen in projection onto the plane of the sky:
\begin{equation}
r_{\rm m} = \frac{\int_0^{r_{\text{out}}} \int_0^{2 \pi} \, I(\rho,\theta) \, \rho^{2} \,\text{d} \theta \, \text{d} \rho}{\int_0^{r_{\text{out}}} \int_0^{2\pi} \, I(\rho,\theta)  \, \rho \,\text{d} \theta \, \text{d} \rho} \hspace{2mm} ,
\label{eq:mean_em_radius}
\end{equation}
where $(\rho,\theta)$ are the polar coordinates in the plane of the sky, and $I$ is the intensity of the emission from the BLR integrated over the spectral dimension\footnote{Eq.~\ref{eq:mean_em_radius} corrects Eq.~8 given in \citet{2017Braibant}.}. The half-light radius  $r_{\rm 1/2}$ is defined by the equation
\begin{equation}
\frac{\int_0^{r_{1/2}} \int_0^{2 \pi} \, I(\rho,\theta) \, \rho \,\text{d} \theta \, \text{d} \rho}{\int_0^{r_{\text{out}}} \int_0^{2\pi} \, I(\rho,\theta)  \, \rho \,\text{d} \theta \, \text{d} \rho} = \frac{1}{2} \hspace{2mm} .
\label{eq:mean_hl_radius}
\end{equation}
The values of $r_{\rm m}$ and $r_{1/2}$ are given in Table~\ref{tab:mehlr}  in units of the BLR inner radius $r_{\text{in}}$ for the different models.

\begin{table}[h]
\caption{BLR mean and half-light radii for the different models}
\label{tab:mehlr}
\centering
\begin{tabular}{lcccc}
\hline\hline
    Model      & $i \;\; (\degr)$ & $q$ & $r_{\rm m} / r_{\text{in}}$ & $r_{1/2} / r_{\text{in}}$   \\
\hline
 KD   &    22   &   1.5   &   4.55   &    4.06    \\
 KD   &    34   &   1.5   &   4.32   &    3.86    \\
 KD   &    44   &   1.5   &   4.10   &    3.64    \\
 KD   &    62   &   1.5   &   3.60   &    3.08    \\
 PW   &    22   &   1.5   &   4.83   &    4.60    \\
 PW   &    34   &   1.5   &   4.76   &    4.38    \\
 PW   &    44   &   1.5   &   4.71   &    4.40    \\
 PW   &    62   &   1.5   &   4.80   &    4.56    \\
 EW   &    22   &   1.5   &   5.89   &    5.96    \\
 EW   &    34   &   1.5   &   5.60   &    5.62    \\
 EW   &    44   &   1.5   &   5.32   &    5.26    \\
 EW   &    62   &   1.5   &   4.69   &    4.42    \\
 KD   &    22   &   3.0   &   2.47   &    1.72    \\
 KD   &    34   &   3.0   &   2.35   &    1.64    \\
 KD   &    44   &   3.0   &   2.23   &    1.56    \\
 KD   &    62   &   3.0   &   1.96   &    1.38    \\
 PW   &    22   &   3.0   &   3.06   &    2.36    \\
 PW   &    34   &   3.0   &   3.01   &    2.24    \\
 PW   &    44   &   3.0   &   2.98   &    2.16    \\
 PW   &    62   &   3.0   &   3.04   &    2.22    \\
 EW   &    22   &   3.0   &   3.73   &    2.94    \\
 EW   &    34   &   3.0   &   3.54   &    2.78    \\
 EW   &    44   &   3.0   &   3.37   &    2.64    \\
 EW   &    62   &   3.0   &   2.97   &    2.26    \\
\hline
\end{tabular}
\end{table}

\end{appendix}

\end{document}